\newif\ifarxiv\arxivtrue

\arxivfalse    % LETTER VERSION

\ifarxiv
\documentclass[12pt,a4paper]{article}
\usepackage{amsthm}
\usepackage{lmodern}
\usepackage[T1]{fontenc}
\usepackage{amsmath,amssymb,bbm} 
\usepackage{accents}
\usepackage{color}
\usepackage{graphicx}
\usepackage[nosort]{cite}
\usepackage[bulletsep]{collref}
\usepackage{tensor}
\usepackage{color}

\graphicspath{{./figures/}}

\usepackage[a4paper,text={173mm,216mm},centering]{geometry}
\makeatletter \let\@keywords\@empty \let\@subject\@empty
\providecommand{\keywords}[1]{\gdef\@keywords{#1}}
\providecommand{\subject}[1]{\gdef\@subject{#1}}
\def\thetitle{\@title}
\def\theauthor{\@author}
\def\thesubject{\@subject}
\def\thedate{\@date}
\def\thekeywords{\@keywords}
\makeatother
\AtBeginDocument{%
\hypersetup{pdftitle={\thetitle}}%
\hypersetup{pdfauthor={\theauthor}}%
\hypersetup{pdfsubject={\thesubject}}%
\hypersetup{pdfkeywords={\thekeywords}}%
}

\providecommand{\hypersetup}[1]{}

\hypersetup{plainpages=false}
\hypersetup{pdfpagemode=UseNone}
\hypersetup{bookmarksnumbered=true}
\hypersetup{pdfstartview=FitH} % open with fit page
\hypersetup{colorlinks=false}
\hypersetup{citebordercolor={.5 1 .5}}
\hypersetup{urlbordercolor={.5 1 1}}
\hypersetup{linkbordercolor={1 .7 .7}}
\allowdisplaybreaks[3]

\usepackage[font=small,labelfont=bf,width=0.85\textwidth]{caption}

\else

\documentclass[amsmath,amssymb,aps,letterpaper,prl,twocolumn,preprintnumbers,showpacs]{revtex4}

\usepackage{graphicx}
\fi

\usepackage{mathtools}% for mathclap

\usepackage{color}

\DeclareFontFamily{U}{mathx}{\hyphenchar\font45}
\DeclareFontShape{U}{mathx}{m}{n}{
      <5> <6> <7> <8> <9> <10>
      <10.95> <12> <14.4> <17.28> <20.74> <24.88>
      mathx10
      }{}
\DeclareSymbolFont{mathx}{U}{mathx}{m}{n}
\DeclareFontSubstitution{U}{mathx}{m}{n}
\DeclareMathAccent{\widecheck}{0}{mathx}{"71}

\makeatletter \let\@keywords\@empty \let\@subject\@empty
\providecommand{\keywords}[1]{\gdef\@keywords{#1}}
\providecommand{\subject}[1]{\gdef\@subject{#1}}
\def\thetitle{\@title}
\def\theauthor{\@author}
\def\thesubject{\@subject}
\def\thedate{\@date}
\def\thekeywords{\@keywords}
\makeatother
\AtBeginDocument{%
\hypersetup{pdftitle={\thetitle}}%
\hypersetup{pdfauthor={\theauthor}}%
\hypersetup{pdfsubject={\thesubject}}%
\hypersetup{pdfkeywords={\thekeywords}}%
}

\usepackage[bookmarks=true,hyperfigures=true]{hyperref}
\usepackage{fixmath}% makes latex math comply with international standards ISO 31-0:1992 to ISO 31-13:1992 (for capital greek letters)

\providecommand{\href}[2]{#2}

\let\oldbfseries=\bfseries
\let\oldmdseries=\mdseries
\let\oldnormalfont=\normalfont
\renewcommand{\bfseries}{\oldbfseries\boldmath}
\renewcommand{\mdseries}{\oldmdseries\unboldmath}
\renewcommand{\normalfont}{\oldnormalfont\unboldmath}

\makeatletter
\newlength{\apb@width}
\newcommand{\autoparbox}[2][c]{\settowidth{\apb@width}{#2}\parbox[#1]{\apb@width}{#2}}

\makeatother

\newcommand{\sfrac}[2]{{\textstyle\frac{#1}{#2}}}

\newcommand*{\diff}{{\mathrm d}}

\newcommand{\Complex}{\mathbb{C}}

\newcommand{\tr}{\mathop{\mathrm{tr}}}
\newcommand{\alg}[1]{\mathfrak{#1}}
\newcommand{\grp}[1]{\mathrm{#1}}

\newcommand{\levz}{\mathrm{J}}

\newcommand{\chargeC}{\mathbb{J}}
\newcommand{\master}{%
 \widehat{\vphantom{{\raisebox{-0.6pt}{$\delta$}}} \smash[t]{\delta}}{} %
}
\newcommand{\boost}{%
 \widecheck{\vphantom{{\raisebox{-0.4pt}{$\delta$}}} \smash[t]{\delta}}{} %
}

\newcommand{\gtilde}{g_u}
\newcommand{\masterWL}{{\chargeC}_\text{\tiny WL}}
\newcommand{\levoWL}{\levz^{(1)}_\text{\tiny WL}}

\title{Master Symmetry for Holographic Wilson Loops}
\begin{document}
%
%%%%%%%%%%%%%%%%%%%%%%%%%%%%%%%%%%%%%%%%%%%%%%%%%%%%%%%%%%%%%%%%%%%%%%%%%%%%%%%%
%%%%%%%%%%%%%%%%%%%%%%%%%%%%%%%%%%%%%%%%%%%%%%%%%%%%%%%%%%%%%%%%%%%%%%%%%%%%%%%%
% ARXIV TITLE PAGE
\ifarxiv
\pdfbookmark[1]{Title Page}{title}

\thispagestyle{empty}

\begingroup\raggedleft\footnotesize\ttfamily
HU-EP-16/17\\
\vspace{15mm}
\endgroup

\begin{center}

\begingroup\Large\bfseries\thetitle\par\endgroup
\vspace{15mm}

\begingroup\large\scshape
Thomas Klose, Florian Loebbert, Hagen M\"unkler \par
\endgroup

\vspace{5mm}

\begingroup\itshape
Institut f\"ur Physik and IRIS Adlershof,\\ 
Humboldt-Universit\"at zu Berlin, \\
Zum Gro{\ss}en Windkanal 6, D-12489 Berlin. Germany
\vspace{3mm}

\par\endgroup

{\ttfamily
\href{mailto:loebbert@physik.hu-berlin.de}{loebbert@physik.hu-berlin.de}
}
\vspace{10mm}

\textbf{Abstract}

\bigskip

\begin{minipage}{13cm}

\end{minipage}

\end{center}

\setcounter{page}{0}

\newpage
%%%%%%%%%%%%%%%%%%%%%%%%%%%%%%%%%%%%%%%%%%%%%%%%%%%%%%%%%%%%%%%%%%%%%%%%%%%%%%%%
%%%%%%%%%%%%%%%%%%%%%%%%%%%%%%%%%%%%%%%%%%%%%%%%%%%%%%%%%%%%%%%%%%%%%%%%%%%%%%%%
% LETTER TITLE PAGE
\else

\title{Master Symmetry for Holographic Wilson Loops}

\preprint{HU-EP-16/17}

\author{Thomas Klose}%
 \email{thklose@physik.hu-berlin.de}
\author{Florian Loebbert}%
 \email{loebbert@physik.hu-berlin.de}
 \author{Hagen M\"unkler}%
 \email{muenkler@physik.hu-berlin.de}
\affiliation{%
Institut f\"ur Physik and IRIS Adlershof,
Humboldt-Universit\"at zu Berlin, \\
Zum Gro{\ss}en Windkanal 6, D-12489 Berlin, Germany
}%

%\date{\today}% 

\begin{abstract}
We identify the symmetry underlying the recently observed spectral-parameter transformations of holographic Wilson loops alias minimal surfaces in AdS/CFT. The generator of this nonlocal symmetry is shown to furnish a raising operator on the classical Yangian-type charges of symmetric coset models. We explicitly demonstrate how this master symmetry acts on strong-coupling Wilson loops and indicate a possible extension to arbitrary coupling.
\end{abstract}


\maketitle

\fi

%%%%%%%%%%%%%%%%%%%%%%%%%%%%%%%%%%%%%%%%%%%%%%%%%%%%%%%%%%%%%%%%%%%%%%%%%%%%%%%%

The AdS/CFT correspondence furnishes a treasure trove for mathematical structures, novel interrelations and physical applications. In particular, its planar integrability has inspired a multitude of innovative methods and is in large part responsible for the progress made towards confirming Maldacena's conjecture. The flagship duality between strings on $\mathrm{AdS}_5\times \mathrm{S}^5$ and four-dimensional $\mathcal{N}=4$ super Yang--Mills theory represents the best understood example of this class of dual theories. The mathematical structure underlying its integrability is Drinfel'd's infinite-dimensional Yangian algebra. This nonlocal symmetry is a typical feature of integrable systems in two dimensions and lies at the heart of string integrability at strong coupling of the above duality \cite{Bena:2003wd,Hatsuda:2004it}. Remarkably, Yangian symmetry has also been discovered for various quantities on the gauge theory side, including the dilatation operator \cite{Dolan:2003uh}, scattering amplitudes \cite{Drummond:2009fd} and Wilson loops \cite{Muller:2013rta,Beisert:2015uda}. In this letter, we add another member to this family of nonlocal symmetries in $\mathrm{AdS}/\mathrm{CFT}$. We demonstrate that symmetric coset models allow for a nonlocal master symmetry $\master$ that captures the models' integrability and has an intriguing set of features: it generates the spectral parameter and acts as a level-raising operator on the Yangian algebra. Mapping conserved charges to conserved charges, $\master$ thus exhibits the characteristic feature of a master symmetry in integrable models. We establish that this nonlocal symmetry underlies the recently observed spectral-parameter deformations of holographic Wilson loops \cite{Ishizeki:2011bf,Kruczenski:2013bsa,Kruczenski:2014bla,Dekel:2015bla,Huang:2016atz}.
With an algebraic understanding of the symmetry at hand, we may hope to formulate a corresponding master symmetry at weak or arbitrary coupling.

\paragraph{Symmetric cosets.}
In the context of the AdS/CFT correspondence, we are primarily interested in strings in the hyperbolic space $\mathrm{AdS}_5$, which can be identified with the coset $\grp{SO}(1,5)/\grp{SO}(5)$. However, our findings apply to arbitrary symmetric spaces $\mathrm{M}=\mathrm{G}/\mathrm{H}$ such that we can keep the discussion general. The space $\mathrm{M}$ is symmetric, if the associated Lie algebras $\alg{g}$ and $\alg{h}$ obey
$\mathfrak{g}=\mathfrak{h}\oplus\mathfrak{m}$, such that
\begin{align}
\left[\mathfrak{h} \, , \, \mathfrak{h} \right] \subset \mathfrak{h} \, , \qquad \left[\mathfrak{h} \, , \, \mathfrak{m} \right] \subset \mathfrak{m} \, , \qquad \left[\mathfrak{m} \, , \, \mathfrak{m} \right] \subset \mathfrak{h} \, .
\end{align}
The dynamical variable is the field $g(z)\in\grp{G}$ with $z$ denoting the complex worldsheet coordinate. To write down the action of the model, we define the flat $\alg{g}$-valued Maurer--Cartan form 
\begin{align}\label{eq:Maurer}
U&=g^{-1}\diff g,
&
\diff U+U\wedge U&=0.
\end{align}
With $U = A + a$, where $A$ and $a$ are projections of $U$ onto $\mathfrak{h}$ and $\mathfrak{m}$ respectively, we can write the action as
\begin{equation}\label{eq:action}
S= \int \tr \left( a \wedge \ast a \right).
\end{equation}
The global $\alg{g}$-symmetry is realized as the infinitesimal transformation $\delta_\epsilon g = \epsilon g$ with $\epsilon\in \alg{g}$. The conservation law for the Noether current associated with this symmetry,
\begin{equation}
j=-2ga g^{-1},
\end{equation}
is equivalent to the equations of motion for $g$. This Noether current is also flat and hence provides the basis for the framework of integrability. The conservation equation and flatness conditions for $j$ are then usually packaged into the condition that the Lax connection
\begin{align}
\ell_u&= \frac{u}{1+u^2} (u\, j +\ast j)\, ,
&
\diff\ell_u+\ell_u\wedge\ell_u&=0 \, ,
\end{align}
is flat for arbitrary values of the so-called spectral parameter $u\in\Complex$. This Lax connection defines a flat deformation $L_u$ of the Maurer--Cartan form through the transformation
\begin{align}\label{eq:Lu}
L_u&=U+g^{-1}\ell_ug ,
&
\diff L_u+L_u\wedge L _u&=0,
\end{align}
and we have $L_0=U$ since $\ell_0=0$.

\paragraph{Master symmetry.}
We now lift the spectral parameter from the level of Lax connections to the physical field $g$. A deformation $\gtilde$ of $g$ can be defined by the following auxiliary linear problem:
\begin{align}
\diff \gtilde&=\gtilde L_u,
&
\gtilde(z_0)&=g(z_0), \label{eq:inival}
\end{align}
with $z_0$ an arbitrary reference point on the worldsheet. Due to the flatness of $L_u$, this initial value problem is well-defined and has a unique solution for simply connected worldsheets. With the ansatz
\begin{align}\label{eq:defgtilde}
\gtilde(z)&=\chi_u(z) g(z),
&
\chi_u(z_0)&=\mathbf{1},
\end{align}
the defining equation \eqref{eq:inival} for $\gtilde$ is satisfied, provided that $\chi_u$ is a solution of
\begin{equation}\label{eq:diffchi}
\diff \chi_u=\chi_u\ell_u.
\end{equation}
Significantly, the transformation $g\mapsto \gtilde$ preserves the action and the equations of motion and hence represents a symmetry. This was observed in~\cite{Eichenherr:1979ci} and will be demonstrated below. Via the definitions \eqref{eq:defgtilde} and \eqref{eq:diffchi} we have thus shown that the spectral-parameter dependence can be translated from the Lax connections into a deformation of the field~$g$. In order for $g_u$ to represent a physical solution of the considered model, we need to restrict $u$ to real values. 
Expanding \eqref{eq:diffchi} about $u=0$, we find the action of the (nonlocal) infinitesimal generator $\master$ as
\begin{align}\label{eq:defmaster}
\master g(z)&=\chi^{(0)}(z) g(z),
&
\chi^{(0)}(z)&= \int_{z_0}^z \ast j.
\end{align}
On the projections of the Maurer--Cartan form \eqref{eq:Maurer}, we thus have
\begin{align}\label{eq:masterona}
\master A&=0,
&
\master a&=-2\ast a.
\end{align}
This directly shows that the equations of motion 
\begin{equation}
\diff \ast a + \ast a \wedge A + A \wedge \ast a = 0
\end{equation}
are invariant under the variation $\master$ due to the condition $\diff a + a \wedge A +  A \wedge a = 0$, which follows from the flatness of $U$.
In fact, we may expand $\chi_u$ in a power series of $u$
\begin{align} 
\chi_u & = \sum_{n=0}^\infty \chi^{(n-1)} u^n,
&
\chi_u^{(-1)}&=1,
\end{align}
which generates a whole tower of master symmetries 
\begin{align}\label{eq:masteru}
\master_u\,g &= \chi_u^{-1} \frac{\diff}{\diff u}\chi_u \, g, 
& 
\master_u &= \sum \limits _{n=0} ^\infty u^n \, \master ^{(n)} ,
\end{align}
as will be argued below.

Let us pause here to indicate various connections to and among the previous literature. The functions $\chi^{(n)}$, which we introduced to mediate between the undeformed and the deformed solution \eqref{eq:defgtilde}, feature in the classical work of Br{\'e}zin, Itzykson, Zinn-Justin and Zuber (BIZZ) \cite{Brezin:1979am} as auxiliary potentials for the construction of conserved nonlocal currents. The BIZZ procedure is rather universal, since it applies to any model possessing a flat and conserved current and does not require the potentials $\chi^{(n)}$ to be generators of a symmetry. 
The large transformation \eqref{eq:defgtilde} was studied in the context of generic symmetric space models by Eichenherr and Forger, who referred to it as a dual transformation that rotates a one-form into its Hodge dual~\cite{Eichenherr:1979ci}, cf.\ \eqref{eq:masterona}. Employing similar ideas, Schwarz provided an extensive analysis of nonlocal symmetries of symmetric space models \cite{Schwarz:1995td}. 
He discussed the interesting Virasoro-like properties of the tower of symmetries $\master^{(n>0)}$ to which we add the master symmetry $\master=\master^{(0)}$.
More recently,
Ishizeki, Kruczenski and Ziama observed that via equation \eqref{eq:inival}
the spectral-parameter deformation induces a familiy of minimal surface solutions in $\mathrm{AdS}_3$  \cite{Ishizeki:2011bf}. Here we identify the underlying nonlocal symmetry transformation and its relation to the Yangian symmetries of the model.
 In  \cite{Beisert:2012ue} Beisert and L\"ucker gave a prescription for the construction of flat Lax connections, which also rotates the components of the Maurer--Cartan form into each other. Their method starts with an operator acting on a flat connection instead of the physical field $g$, but was shown to extend to a multitude of integrable theories of rational type.

\paragraph{Classical Yangian.}
We now show that the integrability of symmetric space models can be captured completely in terms of the global Lie algebra symmetry $\delta_\epsilon$ and the master symmetry generated by $\master$. A general criterion for a variation $\delta g=\eta g$ to be a symmetry of the model is given by \cite{us:2016}
\begin{equation}
g^{-1} \diff \ast ( \diff \eta +[ j , \eta] ) g \in \alg{h}.
\end{equation}
This criterion at hand, it is not hard to show that if $\delta_0$ generates a symmetry, then so does its conjugation with~$\chi_u$:
\begin{equation}\label{eq:conjugation}
\delta_{0,u}\,g=\chi_u^{-1}\delta_0 (\chi_u g).
\end{equation}
Hence, the above master symmetry allows us to turn a symmetry $\delta_0$ into a one-parameter family of symmetries. Applying this procedure to $\delta_0=\delta_\epsilon$ yields the tower of nonlocal symmetries of Yangian-type:
\begin{align} \label{eq:eps,u}
\delta_{\epsilon,u}\,g&=\chi_u ^{-1} \epsilon \chi_u \, g,
&
\delta_{\epsilon,u}=\sum_{n=0}^\infty u^n \,\delta_\epsilon^{(n)}.
\end{align}
The associated conserved currents can be obtained by iterative application of $\master$ to the flat Noether current~$j$. At subleading order we find
\begin{equation}
  \widehat\delta j = -2 \ast j + [\chi^{(0)}, j ],
\end{equation}
which yields the standard expressions for the Yangian level-zero and level-one charges:
\begin{align}\label{eq:Yangcharges}
\levz^{(0)}&=\int * j,
&
\levz^{(1)}&=2 \int j +\int \limits_{\sigma_1 < \sigma_2} [\ast j_1, \ast j_2 ] .
\end{align}
The components of the Noether current $j$ obey a Poisson algebra that is similar to the case of the principal chiral model, for which it was explicitly shown that the resulting charges \eqref{eq:Yangcharges} span a classical Yangian algebra \cite{MacKay:1992he}. This can be extended to the case of symmetric space models at hand \cite{us:2016}. A closed expression for the tower of nonlocal currents associated to the one-parameter family of transformations \eqref{eq:eps,u} is given by
\begin{equation}
j_u = \chi_u \big[ j - \sfrac{2u}{1-u^2} \, \ast j \big] \chi_u ^{-1}, 
\end{equation}
which obeys the relation
$
\master \, j_u = \left( 1 + u^2 \right) \frac{\diff}{\diff u} \,j_u \,.
$
It can be shown that the variation $\master_u$ defined in \eqref{eq:masteru} may be re-expressed as 
\begin{equation}
\master_u \, g = \frac{1}{1+u^2}\,\chi_u^{-1} \,\master (\chi_u \, g),
\end{equation}
i.e.\ in the form \eqref{eq:conjugation}. This shows that $\master_u$ indeed provides a one-parameter family of symmetries. 
\begin{table}
\renewcommand{\arraystretch}{1.4}
\begin{tabular}{| l | l |}\hline
Level-0 Symmetry&Integrable Completion\\\hline\hline
Lie Algebra $\delta_\epsilon=\delta_\epsilon^{(0)}$& Higher Yangian $\widehat\delta_\epsilon^{(n)}$\\\hline
 Master $\master=\master^{(0)}$&  Higher Master $\widehat\delta^{(n)}$\\\hline
 Spacetime& Spacetime $+$ Gauge\\\hline
\end{tabular}
\caption{Overview of basic (level-0) symmetries and their integrable completions using the master symmetry.}% 
\label{tab:towers}
\end{table}%

In order to see that the master symmetry $\widehat \delta$ acts as a raising operator on the Yangian charges, we introduce the real parameter $\theta$ via the relation
\begin{equation}
e^{i\theta} = \frac{1-iu}{1+iu}.
\end{equation}
We then write the different levels of Yangian generators as the coefficients in the expansion
\begin{align}
\levz(\theta) &= \sum \limits _{n=0} ^{\infty} \frac{\theta^n}{n!}\,\levz ^{(n)} \, ,
&
\levz(\theta) &= \int \ast\, j_{u(\theta)},
\end{align}
which implies the raising relation
\begin{align}
\master\, \levz^{(n)}&=\levz^{(n+1)}, 
&
\frac{\diff }{\diff \theta}\levz(\theta)&=\master \,\levz(\theta).
\end{align}
Hence, starting from the Lie algebra charge $\levz$, application of $\master$ yields the infinite tower of Yangian charges. 

Note that the Yangian algebra comes with a natural \emph{lowering} operator defined via $\boost\,\levz^{(1)}=\levz^{(0)}$. The automorphism $\boost$ was introduced by Drinfel'd as a generator of the spectral parameter \cite{Drinfeld:1985rx} and is typically realized by the Lorentz boost in two-dimensional quantum field theory~\cite{Bernard:1990jw}. It is an open problem, however, whether a classical realization of this lowering operator exists~\cite{MacKay:1992he}, which underlines the algebraic interest in the above raising operator~$\master$.

\paragraph{Nonlocal Casimir.}
Since $\master$ was defined as an on-shell symmetry, the Noether procedure is in principle not applicable. However, we are still able to derive associated conserved currents using the equations of motion only in the form $\diff \chi_u=\chi_u \ell_u$. If an off-shell extension of the above symmetries exists, the current we derive will agree with the respective Noether current on shell. Similar comments apply to the Yangian symmetries as introduced above. In the case of principal chiral models it was shown that the latter can be extended to off-shell symmetries \cite{Dolan:1980kz,Hou:1981hn}.  For the master symmetry we formally introduce a free, local parameter $\rho(z)$ into the variation \eqref{eq:defmaster}, i.e.\ we write $\master g = \rho \chi^{(0)} g$, whose application to the action yields
\begin{equation}
\master S=\int \tr ( \ast j \chi^{(0)} ) \wedge \diff \rho \; .
\end{equation}
Here it was used that $\diff \chi ^{(0)} =  \ast j$. Hence, we find that the current $\widehat j=\tr (j \chi^{(0)} )$ is conserved and yields the conserved charge
\begin{equation}
\chargeC := \int \ast \widehat j = \sfrac{1}{2} \tr(\levz \levz) \, .
\end{equation}
This is the quadratic Casimir of the Lie algebra charge $\levz$ and it comes as no surprise that it is conserved. However, it is interesting that a symmetry $\widehat \delta$ exists, which yields the quadratic Casimir as an associated conserved charge. 

In analogy to the tower of Yangian generators, we may obtain a tower of Casimir charges by iterative application of $\widehat\delta$ to $\chargeC$:
\begin{align}\label{eq:towerCascharges}
\chargeC (\theta) &= \sfrac{1}{2}  \tr\big(\levz(\theta) \levz(\theta)\big)  \, ,
&
\master \,\chargeC (\theta)&= \frac{\diff}{\diff \theta}\chargeC (\theta).
\end{align}
At subleading order we find $\chargeC ^{(1)} = \tr \big( \levz \, \levz ^{(1)} \big)$, which can easily  be  checked to Poisson-commute with the Lie algebra generators $\levz$. It is worth noting that the Casimir $\mathbb{J}$ does not generate the master symmetry $\master$ via its Poisson bracket with the field $g$. The same holds for the nonlocal Yangian symmetry which is generated via a Lie--Poisson action and not by the ordinary symplectic action \cite{Babelon:1991ah}.

It turns out that the conjugation of the basic spacetime symmetries (translations, boost, conformal transformations) with $\chi_u$ does not result in new transformations but merely yields a $u$-dependent linear combination of spacetime and gauge symmetries, cf.\ Table \ref{tab:towers}.

\paragraph{Algebra relations.}
An obvious aspect of interest is the algebra of the above nonlocal symmetries. We find the following commutators between the master and the Yangian variations $\master ^{(n)}$ and $\delta_\epsilon^{(n)}$, respectively:
\begin{align}
\left[ \delta_{\epsilon_1}^{(n)} , \delta_{\epsilon_2}^{(m)} \right]
&=
\delta_{[\epsilon_1,\epsilon_2]}^{(m+n)} + (-1)^n a_{n,m} \, \delta_{[\epsilon_2 , \epsilon_1 ^\prime ]} ^{(m-n)}  \nonumber
\\
\left[ \master^{(n)} , \master^{(m)} \right] 
&=
(n-m) \, \master ^{(n+m+1)} +  \master^{(n-m-1)}_m - \master ^{(m-n-1)}_n 
\nonumber\\
\left[ \master^{(n)} , \delta_{\epsilon}^{(m)} \right] 
&= 
- m\big[ \delta_{\epsilon} ^{(m+n+1)} + (-1)^n \delta_{\epsilon} ^{(m-n-1)}   \big].
\end{align}
Here, we defined 
$a_{n,m}=(1-\delta_{m,0})(1-\delta_{n,0})$
 as well as the shorthand notations
$\master^{(n)}_m=(n+2m+3)(-1)^m \master^{(n)}$, 
 $\delta_\epsilon ^{(-n)} = (-1)^n \delta_{\epsilon ^\prime} ^{(n)}$ and $\master^{(-n)}=0$. We furthermore obtain
the expression $\epsilon^\prime =  \epsilon|_\alg{h}-\epsilon|_\alg{m}$, if 
we set $g ( z_0 ) = \mathbf{1} $.
Except for the master symmetry $\master=\master^{(0)}$, the commutators of similar generators were considered by Schwarz~\cite{Schwarz:1995td}, who argues that the~$\master^{(n>0)}$ relate to a centerless Virasoro-like algebra. 
This relation to the Virasoro algebra is reminiscent of the Sugawara construction and is a typical feature of master symmetries in integrable models, see e.g.\ \cite{Moerbeke,2000CMaPh.211...85Z}. Note that the algebra of variations is not expected to coincide with the Poisson algebra of charges since nonlocal symmetries are typically generated in a nonlinear way \cite{Babelon:1991ah}.

\paragraph{Wilson loops.}
A beautiful arena for the application of the above master symmetry is provided by the AdS/CFT correspondence \cite{Maldacena:1997re}. In particular, this duality relates the expectation value of the Maldacena--Wilson loop in $\mathcal{N}=4$ super Yang--Mills theory at strong coupling to a minimal surface in $\mathrm{AdS}_5$ ending on the loop contour $\gamma$ \cite{Maldacena:1998im,Rey:1998ik}. Specifically the expectation value reads
\begin{align}
W(\gamma) \overset{\lambda \gg 1}{=} 
\exp{ \Big[ - \sfrac{\sqrt{\lambda}}{2 \pi} A_{\mathrm{ren}}(\gamma) \Big] } \, .
\end{align}
Here, $A_{\mathrm{ren}}(\gamma)$ denotes the appropriately renormalized area of a minimal surface ending on the boundary contour~$\gamma$. The area functional is given by a string action in $\mathrm{AdS}_5$ and since $\mathrm{AdS}_5$ is a symmetric space, the above setup applies. In particular, we consider the master symmetry $\master$ as well as the associated large transformation $g \mapsto g_u$ and derive the variation of a generic boundary curve $\gamma$ under this symmetry. The derivation shows that the transformation maps the conformal boundary to itself, which is essential for the possibility to restrict the transformation to the class of holographic Wilson loops. Moreover, one can show that the transformation $g \mapsto g_u$ is not only a formal symmetry of the area functional, but also a symmetry of the renormalized area of the minimal surface, i.e.\ one has $A_{\mathrm{ren}}(\gamma) = A_{\mathrm{ren}}(\gamma_u)$ \cite{us:2016}.

In order to discuss the above transformations for minimal surfaces in $\mathrm{AdS}_5 \simeq \grp{SO}(1,5)/\grp{SO}(5)$ in some detail, we employ the standard generators $M_{IJ}$ of $\grp{SO}(1,5)$. These generators obey
\begin{align}\label{eq:comms}
\left[ M_{IJ} , M_{KL} \right] &= \eta_{I [ K} M_{L] J} - \eta_{J [ K} M_{L] I} \, , 
\end{align} 
as well as $\tr \left( M_{IJ}  M_{KL} \right) = 2 \eta_{I [L} \eta_{K] J}$. The gauge algebra $\alg{h}=\alg{so}(5)$ is spanned by $M_{\mu\nu}$ and $M_{\mu 5}$ for $\mu,\nu=1,\ldots,4$. In order to introduce Poincar\'e coordinates $(X^\mu,y)$ on the coset space, we define the combinations
\begin{align*}
P_\mu = M_{\mu 6} - M_{\mu 5} \, , \quad K_\mu = M_{\mu 6} + M_{\mu 5} \, , \quad D = M_{56} \, ,
\end{align*}
and write the coset representatives as $g = e^{X \cdot P} y^D$.
A variation $\delta g$ is then linked to the variations of the coordinates $\delta X^\mu$ and $\delta y$
by
\begin{equation}
g^{-1} \delta g - h = \frac{\delta X^\mu}{y} P_\mu + \frac{\delta y}{y} D ,
\end{equation}
where $h(z)$ is an element of $\alg{h}$ accounting for possible gauge transformations.
In the case of $\mathrm{AdS}$-isometries $\delta_a g = T_a g$, where $T_a$ denotes the basis elements of $\alg{g}$, the variation of the boundary curve coordinates $x^\mu(\sigma)$ is given by $\delta_a x^\mu(\sigma) = \xi^\mu_a(x(\sigma))$. Here, the $\xi^\mu_a$ form a basis of conformal Killing vectors of the flat boundary space, see e.g.\ \cite{Munkler:2015gja}. For the master symmetry, we employ the known form of the minimal surface solution close to the boundary \cite{Polyakov:2000jg} to obtain the variations $\master y = \mathcal{O}(y)$ and
\begin{equation}\label{eq:genxmu}
\master x^\mu (\sigma) = - G^{ab}  \int  _0 ^\sigma \diff \sigma_1 \, \frac{\delta A_{\mathrm{ren}}(\gamma)}{\delta x^\nu_1 } \,  \xi^\nu _a (x_1) \, \xi^\mu _b (x(\sigma)) \, ,
\end{equation}
where we introduced $G_{ab} = \tr(T_a T_b)$ to denote the metric on $\alg{g}$ and set $x^\mu_k=x^\mu(\sigma_k)$. The invariance of the minimal surface under the variation $\master$ is then encoded in the fact that the generator
\begin{equation}\label{eq:genstrongWL}
\masterWL =  G^{a b}  \int\limits_{\mathclap{\sigma_1 < \sigma_2}}   \diff \sigma_1   \diff \sigma_2  \,
\xi ^\nu _a (x_1) \, \frac{\delta A_{\mathrm{ren}}(\gamma)}{\delta x^\nu_1 } \,  
\xi ^\mu _b (x_2) \, \frac{\delta}{\delta x_2^\mu } 
\end{equation}
annihilates the renormalized area $A_{\mathrm{ren}}(\gamma)$ of the minimal surface. 
Similarly, one may evaluate the higher master generators $\masterWL^{(n)}$ on Wilson loops yielding conditions which are tightly entangled with the higher levels of Yangian symmetry.
\begin{figure}
\includegraphics[scale=.44]{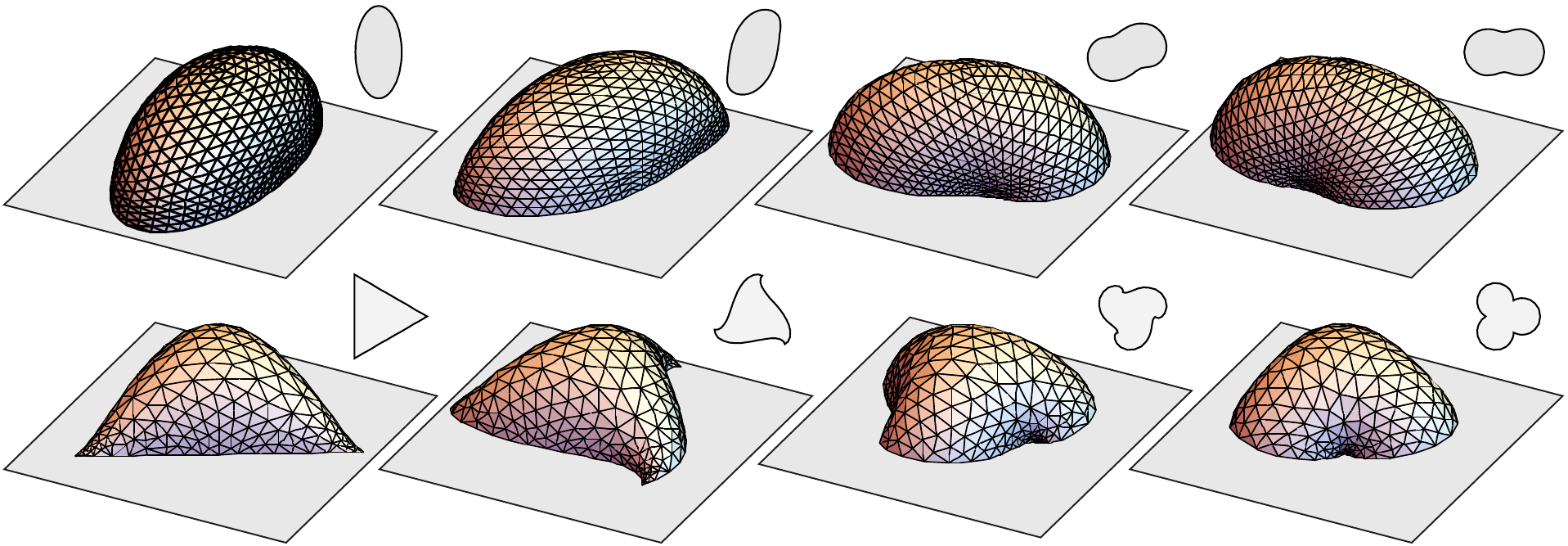}
\caption{Sequence of master transformations for parameter values $\theta = 0, \frac{3\pi}{16}, \frac{3\pi}{4},\pi$ applied to discrete minimal surfaces bounded by an ellipse and a triangle at $y = \frac{1}{10}$.}
\label{fig:sequence}
\end{figure}

An immediate point of interest is the geometric meaning of the spectral-parameter transformation \eqref{eq:genstrongWL} evaluated on Wilson loops. In fact, our analysis shows that studying explicit transformations requires the knowledge of full nontrivial minimal surface solutions, since the information contained in the universal orders of the near-boundary expansion \cite{Polyakov:2000jg} is insufficient to determine the functional derivative term featuring in the variation \eqref{eq:genstrongWL}. This suggests to employ a numerical approach from which we obtain the concrete examples in FIG.\ \ref{fig:sequence}. The details of this numerical evaluation are provided in \cite{us:2016}.

Let us emphasize some consequences on the Yangian symmetry of strongly
coupled Maldacena--Wilson loops. In \cite{Muller:2013rta} the
corresponding generators were set up to yield the respective conserved
charges when evaluated on minimal surfaces. For this it was not necessary
to know the underlying classical symmetry generator on the field~$g$.
The relations established in the present paper show that it is given by $\delta_c ^{(1)}$.
Finding the level-1 variation of the boundary curve is analogous to the above discussion for the master symmetry. The variation 
$\delta_c^{(1)}x^\mu$ is obtained from \eqref{eq:genxmu} by replacing $G^{ab}$ by $f_c{}^{ba}$ and the level-1 generator $\levoWL$ follows from \eqref{eq:genstrongWL} analogously.
Applying this generator to the minimal area gives the Yangian Ward identity for the Maldacena--Wilson loop at strong coupling. The local term entering this identity can now be interpreted as a boundary term arising from the application of $\delta ^{(1)}_c$ to the area functional.  

\paragraph{Beyond strong coupling.}
It is natural to ask whether the above master transformation furnishes a symmetry beyond strong coupling.
At weak coupling, Dekel observed that the expectation value of the Maldacena--Wilson loop is \emph{not} invariant for contours obtained as boundaries of spectral parameter deformed minimal surfaces in $\mathrm{AdS}_3$ \cite{Dekel:2015bla}.
Hence, beyond strong coupling we generically have $\masterWL W (\gamma) \neq 0$, if we naively apply the strong-coupling generator \eqref{eq:genstrongWL}. However, above we have shown that the Wilson-loop deformation corresponds to a symmetry of the underlying model. There is in fact no indication that the representation of the symmetry generator should be independent of the 't~Hooft coupling~$\lambda$. As a generalization of \eqref{eq:genstrongWL}, it is thus natural to consider the following operator for any value of the coupling $\lambda$:
 \begin{equation}\label{eq:genstrongWLgeneral}
\masterWL^{(\lambda)} = \int\limits_{\mathclap{\sigma_1 < \sigma_2}}   \diff \sigma_1   \diff \sigma_2  \,
\xi ^{\nu a}  (x_1) \, \frac{\delta \log W (\gamma) }{\delta x^\nu_1 } \,  
\xi ^\mu _a (x_2) \, \frac{\delta}{\delta x_2^\mu } \, .
\end{equation}
This generator reproduces the transformation \eqref{eq:genstrongWL} at strong coupling and it is easy to see that it annihilates the expectation value $ W (\gamma) $ of the Maldacena--Wilson loop for any $\lambda$. Moreover, one can show that it commutes with the generators of conformal transformations and thus it appears to be an appropriate generalization of the master symmetry on Wilson loops. 
Similarly, the higher variations in the tower of master
symmetries \eqref{eq:masteru} should be applied to Wilson loops yielding nontrivial
constraints as consequences of integrability.

 \paragraph{Outlook.} 
Let us highlight some particularly attractive directions for further investigation. 
An important goal is to better understand the geometric meaning of the above master symmetry using explicit examples of minimal surfaces \cite{us:2016}, cf.\ FIG.\ \ref{fig:sequence}.
Moreover, the string theory underlying the $\mathrm{AdS}_5/\mathrm{CFT}_4$ correspondence is based on a supersymmetric coset model with $\mathbb{Z}_4$-grading. It would be desirable to explicitly work out the above nonlocal symmetries for this case; see \cite{Beisert:2012ue} for the corresponding construction starting from the Maurer--Cartan form. 
This would allow us to make contact to the Wilson loops in superspace introduced in \cite{Beisert:2015jxa}, for which the observed Yangian symmetry at strong coupling \cite{Munkler:2015gja} has a counterpart at weak coupling \cite{Beisert:2015uda}. Finally, it would be highly interesting to identify an analogue of the above master symmetry on the level of the action or the equations of motion of $\mathcal{N}=4$ super Yang--Mills theory.

\paragraph{Acknowledgements.}
We thank  A.~Dekel, H.~Dorn, B.~Hoare, T.~McLoughlin, J.~Plefka, A.~Tseytlin and G.~Yang for useful discussions.
This research is supported in part by the SFB 647 \textit{``Space-Time-Matter. Analytic and Geometric Structures''} and the Research Training Group GK 1504
\textit{``Mass, Spectrum, Symmetry''}.

%%%%%%%%%%%%%%%%%%%%%%%%%%%%%%%%%%%%%%%%%%%%%%%%%%%%%%%%%%%%%%%%%%%%%%%%%%%%%%%%
\bibliography{Boost}

\end{document}